\documentclass[smallextended,final]{svjour3}
\usepackage{graphicx,csquotes,amsmath,amsfonts,amssymb,natbib,xcolor}
\usepackage[american]{babel}
\usepackage[a4paper,margin=3cm]{geometry}

\journalname{Forthcoming in Foundations of Physics}

\begin{document}
\sloppy
\raggedbottom

\title{Whence deep realism for Everettian quantum mechanics?\thanks{Order of authorship does not represent priority; authors contributed equally to this work.}}

\author{Raoni Wohnrath Arroyo \and Jonas R. Becker Arenhart}

\institute{Raoni Wohnrath Arroyo \at Centre for Logic, Epistemology and the History of Science, University of Campinas, Campinas, SP, Brazil. Support: grant \#2021/11381-1, São Paulo Research Foundation (FAPESP)\\ Research Group on Logic and Foundations of Science (CNPq), International Network on Foundations of Quantum Mechanics and Quantum Information, Florianópolis, SC, Brazil \\ \email{rwarroyo@unicamp.br}\\Corresponding author \and Jonas R. Becker Arenhart \at Federal University of Santa Catarina, Department of Philosophy, Florianópolis--SC, Brazil \\Federal University of Maranhão, Graduate Program in Philosophy \\ Research Group on Logic and Foundations of Science (CNPq), International Network on Foundations of Quantum Mechanics and Quantum Information, Florianópolis, SC, Brazil\\ Partially funded by CNPq (National Council for Scientific and Technological Development) \email{jonas.becker2@gmail.com}}

\date{This version of the article has been accepted for publication, after peer review (when applicable) but is not the Version of Record and does not reflect post-acceptance improvements, or any corrections. The Version of Record is available online at: http://dx.doi.org/10.1007/s10701-022-00643-0. Use of this Accepted Version is subject to the publisher's Accepted Manuscript terms of use https://www.springernature.com/gp/open-research/policies/acceptedmanuscript-terms}

\dedication{Received: April 25, 2022 / Revised: August 16, 2022 / Accepted: October 18, 2022}

\maketitle

\begin{abstract}
`Shallow' and `deep' versions of scientific realism may be distinguished as follows: the shallow realist is satisfied with belief in the existence of the posits of our best scientific theories; by contrast, deep realists claim that realism can be legitimate only if such entities are described in metaphysical terms. We argue that this methodological discussion can be fruitfully applied in Everettian quantum mechanics, specifically on the debate concerning the existence of worlds and the recent dispute between Everettian actualism and quantum modal realism. After presenting what is involved in such dispute, we point to a dilemma for realists: either we don't have the available metaphysical tools to answer the deep realist's demands, and realism is not justified in this case, or such demands of metaphysical dressing are not mandatory for scientific realism, and deep versions of realism are not really required. 

\keywords{Everettian actualism \and
Everettian quantum mechanics \and
possible worlds \and
quantum modal realism \and
scientific realism}
\end{abstract}

\section{Introduction}\label{null:sec:1}

As \citet[211]{psillos2012} declares, ``[i]t is still an open question ---to me at least--- exactly how much metaphysics (scientific) realism requires or implies \textelp{}''. The question is indeed an important one, given that it has been recently argued that without an associated \emph{metaphysical} picture, there's no way to clearly spell out what one is being a scientific realist about \citep{chakravartty2007metaphysics, french2014, French2018RealMetaph}, so one cannot legitimately be a scientific realist without delving into deep metaphysical questions. This is `deep' realism, as opposed to `shallow' realism. The requirement of providing this additional metaphysical clothing of the posits of a theory is often called ``Chakravartty's Challenge''. A famous solution to the Challenge is the Toolbox Approach to metaphysics (aka `Viking approach'): check on the philosophical literature for methods and theories that were developed in a completely \textit{a priori} fashion and that can provide one with a suitable metaphysical picture, and plug them onto the posits of scientific theories \citep{french2014, french2018jgps, french2012toolbox, FrenMcken2015toolboxagain}. This short description of the problem already indicates a distinction between ontology and metaphysics based on their subject matter, the former dealing with existence-questions (i.e., the posits of science) and the latter dealing with nature-questions (i.e., what they are and how to classify them, in metaphysical terms) \citep{arenhart2012ontological, Hofweber2016MetOnt, arenhartarroyo2021, arroyoarenhart2021}. Under this distinction, that we shall also adopt thoughout this paper, ontology is given by the existential commitments of a theory, and metaphysics is the additional content required to address the question of the natures of such posits, as required to convert a shallow realism into a deep one. Section \ref{null:sec:2} discusses that.

It is almost common sense to state that the solutions of the quantum measurement problem depend on the ontology of a particular interpretation of quantum mechanics one endorses \citep{ruetsche2015shaky, Ruetsche2018GetRealQM, maudlin2019quantumphil, durrlazarovici2020understandingqm}. In this paper, we adopt this line of reasoning and address the demand for scientific realism as framed by Chakravartty's Challenge concerning Everettian quantum mechanics (EQM), which for the purposes of this paper refers to a general claim concerning how to answer to the measurement problem, dealing with the notion of branching and that started with Hugh Everett ---two versions of which we discuss in this paper. As Ladyman \citet[p.~154]{ladyman2010manyworlds} already stressed out, the debate concerning the existence of multiple worlds (for and against) is one of the central features of the philosophical literature concerning EQM. One of the main reasons for this is that there are \textit{at least} two ontologies within the context of the EQM solution to the measurement problem: the relative facts interpretation (RFI), which is a one-world interpretation \citep{Barrett2011everettoneworld, Conroy2012everettnote} whose ontology is expressed by Conroy's `Everettian actualism' \citep{Conroy2018actualism}; and the many-worlds interpretation (MWI) \citep{wallace2012emergent}, which postulates many worlds and whose ontology is nicely captured by Wilson's quantum modal realism \citep{wilson2020QMmodal}. Shallow realists, being concerned only with ontology, would stop here. But those who assume that Chakravartty's challenge must be addressed can't: they should keep on swimming, looking for the appropriate metaphysics. However, there is no information on the metaphysical profile for the worlds of quantum modal realism or Everettian actualism: the dispute is traditionally focused solely on the \textit{existence} (be it in favor or denial of existence) of worlds ---not on their \textit{nature}. The information on the nature of the worlds seems to be found nowhere in the Toolbox. This is the subject of Section \ref{null:sec:3}.

This situation presents a methodological dilemma: either we don't have an available metaphysical profile (and so we should tailor one if we insist that it is important) or the demand for such a clear picture is not really mandatory. Naturalists about the metaphysics of EQM might expect that the answer is to be found in quantum physics itself \citep{wallace2012emergent}. But even if we pretend that EQM tells us unequivocally whether there are many worlds (which it does not), a physical description of the metaphysical profile won't help: it's just physics again. On the other hand, the metaphysical toolbox is empty on the matters of `\textit{what is the nature of metaphysically possible worlds?}', so instead of metaphysical underdetermination, we have \emph{metaphysical null-determination}: no options whatsoever on this metaphysical account of the worlds.\footnote{As an anonymous referee pointed out, one must be careful with such a claim, mainly because ---at least in MWI--- the worlds are actual, so it seems that ``they are not possible worlds in the sense used e.g. in modal logic''. However, the multiplicity of worlds in the MWI is \textit{one way} of treating possible worlds, viz. to treat them as actual (not merely possible).} This may indicate, then, that taking Chakravartty's Challenge too seriously can be problematic.

The reasoning is simple: if we cannot associate a metaphysical profile to the ontological commitments of EQM (be it those of RFI or MWI), we cannot claim an appropriate scientific-realistic attitude about this very theory \citep{French2018RealMetaph}; i.e., if the only legitimate realism is `deep' realism, metaphysical null-determination tells us that we cannot be realists about EQM yet. This way one ends up judging physics by metaphysical criteria, which calls for epistemic justification inasmuch as metaphysics floats free from physics \citep{arroyoarenhart2021}. Section \ref{null:sec:4} discusses this, and Section \ref{null:sec:5} presents the conclusion.

\section{The Challenge}\label{null:sec:2}

From a practical point of view, non-relativistic quantum mechanics (QM) works. This is undisputed. Now, from a philosophical point of view, it could seem that a natural attitude towards such empirical success is to believe that the theory fulfills also some kind of metaphysical expectation, providing for something like a description of how the world looks in its metaphysical aspects.

However, as the century-long debates concerning scientific realism in the quantum realm attests, this is not straightforward. The worldview offered by science (e.g. the equations of contemporary physics) is very crude, leaving much room for speculation. Some argue that the gap between the scientific description and what the world is really like needs a metaphysical filling \citep{french2014, French2018RealMetaph}. 

The question that arises from the methodological point of view, then, is: how much beyond science should scientific realism go? Put it in another way: to which extent should one's scientific realism be informed by metaphysics? Answering such questions, one must pay attention to the epistemic risk \citep{chakravartty2017scientific}. After all, the further one moves away from science, the greater the epistemic risk, i.e., the risk of complementing science with posits that have no specific role in the explanatory economy of science itself. The greater the distance from the actual working of science, the greater the epistemic risk of idle positing. In what follows, we will adopt the taxonomy offered by \citet{French2018RealMetaph}, which classifies the proposals into two methodological attitudes towards this question: the `shallow' scientific realism and the `deep' one.

Scientific realism of the `deep' type is the methodological attitude that argues that we must go beyond science searching for metaphysical investigations that can be used to \textit{clarify} the scientific theory itself, contributing to a coherent worldview. It is, therefore, an attitude with less epistemic humility, but its epistemic risk increases precisely because of this. On the other hand, the scientific realist of the `shallow' kind would only go as far as science goes in the description of the world.

Perhaps the best example is metaphysical naturalism. Understood in broad strokes, naturalism is the thesis that philosophy is continuous with science. As \citet[pp.~3--4]{wallace2012emergent} nicely characterizes it, naturalism is ``\textelp{} the thesis that we have no better guide to metaphysics than the successful practice of science''. The naturalization of ontology and of metaphysics is a hotly debated topic, and typically it is thought that ontology and/or metaphysics may be derived from the science itself, and not from a supra-scientific source, being endowed thus with some of the epistemic privileges of science. The (alleged) advantage of this is to transfer the epistemic credentials of science to metaphysics, so the epistemic risk of the latter would also be lower, when compared with the epistemic risk of analytic metaphysics developed on a purely \textit{a priori} basis--- recall, the closer to science, the lower the epistemic risk of idle positing of explananda that are not warranted by our best account of reality \citep[see also][]{ArroyoArenhart2022Epistemic}.\footnote{We would like to thank an anonymous referee for pointing that out.} The level of epistemic humility involved in this strategy, however, is very high: if we should learn metaphysical lessons solely by paying attention to science, as sometimes is suggested \citep[cf.][]{maudlin2007}, then really nothing is said beyond what science already says. At the risk of just repeating physics, fully naturalized metaphysics may sound irrelevant for metaphysical purposes. As \citet[3431]{ruetsche2015shaky} states, ``[t]he concern is that \textelp{} accepting the certified results of science, requires us merely to hang around science labs intoning `Yup'.'' That is, if all one can do, from the scientific realist perspective, is to account for the internal truths given by ontological frameworks of science, there seems to be not much work for realists to do \citep[cf.][]{arroyo-dasilva2022}.

This brings us to one of the central points of this article. Inspired by \citet[p.~26]{chakravartty2007metaphysics}, who stated that ``[o]ne cannot fully appreciate what it might mean to be a realist until one has a clear picture of what one is being invited to be a realist about'', \citet{french2014, French2018RealMetaph} puts the search for a metaphysical profile for the posits of science as a necessary condition for an attitude to be considered scientific realist. This is how the so-called ``Chakravartty's Challenge'' arises: the challenge of balancing the need to provide this metaphysical profile for theories, so that metaphysics can clarify physics, with a certain degree of epistemic humility ---neither too high, in order not to fall into the `shallow' part of scientific realism, nor too low, after all, there is no privileged point of view of reality such as a `view from nowhere' from which we can say with total certainty how reality is. \citet[p.~48]{french2014} is emphatic to state that those who deny Chakravartty's Challenge are `ersatz' scientific realists or empiricists in denial.

Perhaps the relationship between the `shallow' and `deep' kinds of scientific realism can be better appreciated if we keep in mind a distinction between the terms `ontology' and `metaphysics' made by differences in their respective subject matter \citep[cf.][]{arenhartarroyo2021veritas}. Before we proceed, let us briefly discuss this. Two distinct questions are typically conflated when it comes to describing reality according to quantum mechanics: the ontological question on the one hand, and the metaphysical question on the other. The latter is more general than the former, and the distinction can be firstly put as follows.\footnote{A detailed defense for such a distinction of philosophical disciplines can be found in \citet{arenhartarroyo2021manu}.} Following \citet[p.~13]{Hofweber2016MetOnt}, ontology is a subset of metaphysics (so we can refer to metaphysics qua ontology and metaphysics qua metaphysics, respectively), so that metaphysics qua metaphysics is the philosophical discipline that pursuits the most general questions on the nature of being (e.g. of free will, of mind, the nature of properties, the principle of individuality, etc.) while metaphysics qua ontology deals with questions of modes of existence (e.g. what exists, what is the kinds of things that exists, what is the scope of the existential quantifier, etc.). In the same vein, \citet[244]{thomsonjones2017existencenature} distinguishes ontology and metaphysics as the kinds of philosophical questions they pose, calling them ``nature questions'' and ``existence questions'', respectively.

Also acknowledging the above distinction between existence questions and nature questions, \citet[16]{psillos2021} states that they refer to levels of description as follows: on the ontological level, there is the question of the things ---and kinds of things--- that exists in the world (e.g. tectonic plates, DNA molecules, electrons, etc.); the metaphysical level goes above, asking the metaphysical categories of the things established by the ontological level (e.g. whether the objects are particulars or universals). He seems, however, sure that one should not eat so much metaphysics pie in order to maintain a scientific realist stance. In particular, \citet[29]{psillos2021} explicitly states that we ``don't have to settle all the ontological issues concerning X in order to be committed to the existence of X''. That said, \citet[p.~150, original emphasis]{psillos1999} also recognizes that ``[i]t is certainly arguable that knowing which laws an entity obeys does not exhaust knowing what this entity \textit{is}''. However, it seems safe to say that this position also exemplifies the maxim according to which physics is metaphysics enough. Otherwise, we are being old-fashioned and outdated philosophers discussing medieval notions of `forms' and `substances', when ``[s]uch talk has been overthrown by the scientific revolution of the seventeenth century'' \citep[149]{psillos1999}.

This amount of metaphysics (qua ontology) that Psillos suggests one to buy into scientific realism may settle the question. The shallow realists can just take the scientists' view of their theories and declare them to be approximately true, while deep realists need to base their realism on a full-fledged ontological model along with a metaphysical description of the corresponding posits. Fleshing out realism, from the perspective of shallow realism, would amount exactly to providing an interpretation and to answering the empirically undecidable questions raised by choosing a specific interpretation.

This is clearly not metaphysics enough to deep realists. To \citet[22]{french2019}, ``\textelp{} realists should not be content with adopting a `shallow' form of their stance, as represented by expressing a belief in the existence of electrons, say, and leaving it at that''. He suggests that scientific realists should go ``deep'' into metaphysical questions (qua metaphysics) and fulfilling the metaphysical (qua metaphysics) clear picture demanded by Chakravartty's Challenge in order to have a legitimate realist stance ---viz., the deep realist stance. Only then one can ``\textelp{} offer a clearer picture of what that belief in electrons consists in'' \citep[22]{french2019}.

Thus, merely repeating the features of the theory is not enough \citep[p.~394]{French2018RealMetaph}. Some ``fleshing out'' ---in metaphysical terms--- of what science tells us is required. One of the most full-blown endeavors to do this kind of work is the \textit{Viking/Toolbox Approach} to the metaphysics of science, as presented by French himself. The Viking/Toolbox approach labors on the assumption that metaphysical concepts, as developed by armchair metaphysical speculation, are in general compatible with science, or with the posits they aim to dress in metaphysical clothes. So one should draw on the tools of analytic metaphysics (qua metaphysics) in order to have a clear picture of the ontological posits of the theory ---and hence adopting a legitimate realist stance towards the theory. In this sense metaphysics is understood as a ``Toolbox'' \citep[see also][]{french2012toolbox, FrenMcken2015toolboxagain}, having theories and tools that can be used by the philosophy of science in the service of science, lending itself to enable a response to Chakravartty's Challenge, and, consequently, a plunge into the deepest waters of metaphysics associated with `deep' scientific realism. Notice that existence questions are being tacitly bracketed into the shallow spectrum of scientific ``realism''. To be sure, the shallow realist is stopping at the existence level: ``\textelp{} accepting that there are electrons, for example, but refusing to go any further and state what sorts of things electrons are'' \citep[395]{French2018RealMetaph}.

Deep realists, then, are even more demanding: not deep enough into metaphysics, not \textit{realist} enough. But besides the not so cheerful labels attached by French, can one advance positive arguments for going deep into metaphysics? As it is argued, a layer of metaphysical explanation is indispensable in order to yield an understanding about one's realist commitments. The notion of `understanding' is thus crucial for this view. Shallow realism, as it is alleged by deep realists, does not provide us with \textit{understanding}.

\begin{quote}
    \textelp{} ontological neutrality and metaphysical minimalism raises concerns as to whether we obtain the clear understanding of how the world is that we associate with scientific realism. \textelp{} the realist cannot rest content with epistemology but must seek an understanding articulated in metaphysical terms. \citep[7]{french2014}.
\end{quote}

To sum up: ontology deals with existence questions, with establishing a catalog of what exists. Metaphysics builds from there, describing in general categories what those entities are, or what is the nature of that what exists \citep[see also][for further discussions on this distinction in the metaphysics of science]{arenhartarroyo2021manu, jonas2019filomena}. Quantum mechanics populates the world with electrons, protons, and other kinds of entities and processes. Metaphysics describes those items in metaphysical terms (using metaphysical conceptual machinery, such as `individual', `universal', `tropes', and so on). Shallow realists deal exclusively with questions of ontology (existence), while the deep realists would be the ones that face questions of metaphysics (nature). And the deep level is motivated through a notion of `understanding', which is said to be untenable without the metaphysical level.

Now let us see how this methodological debate can be applied to recent discussions on EQM.

\section{Shallow and deep realism for EQM}\label{null:sec:3}

According to \citet[p.~154]{ladyman2010manyworlds}, three basic questions must be distinguished when discussing EQM:

\begin{quote}
(1) Is EQM coherent?\\(2) Is EQM the best interpretation of quantum mechanics?\\(3) Are there multiple worlds?
\end{quote}

We will not discuss (1) and (2); instead, we will take EQM at face value and remain as neutral as possible with regard to theoretical underdetermination concerning the other interpretations of QM. A fierce defense of EQM concerning these points can be found in \citet{wallace2012emergent} and references therein. We will focus on (3) only. So let us take a step back: is the existence of worlds somehow \textit{implied} by EQM? The short answer is ``no''. The long answer demands a discussion on EQM.

So, how come the EQM? In EQM, the states of physical systems are represented by vectors in Hilbert space, which evolve in time according to the Schrödinger equation. Think about Schrödinger's cat scenario. After being locked up with a venom flask and a quantum-mechanical device, there are two quantum-mechanical descriptions of possible outcomes for of the state of affairs of the macroscopic cat: $|\psi_1\rangle+|\psi_2\rangle$, being $|\psi_1\rangle$ is the description for the scenario in which the cat dies, and $|\psi_2\rangle$ is the scenario in which it lives. \textit{Some} of those two scenarios will take place after one hour, and $|\psi'\rangle$ describes that. As the $|\psi\rangle$ description is a linear process, the transition from $|\psi\rangle$ to $|\psi'\rangle$ is deterministic, to some non-arbitrary level of probability. The temporal evolution of $|\psi\rangle$ to $|\psi'\rangle$ is represented by the Schrödinger equation, which implies that the state of affairs of the box and its contents, after one hour, is

\begin{equation}
|\psi'\rangle=a|\psi_1\rangle+b|\psi_2\rangle\label{eq:simplesuperp} 
\end{equation}
\color{black}

If the Hamiltonian and the Schrödinger equation description is all that goes on in the physical reality, it is implied that we often find ourselves in superposed states, yet we never feel like being in a superposition of macroscopically distinct states (e.g., seeing a superposition of cat states). This is roughly the measurement problem. To cope with our experience, there are at least three solutions to the measurement problem: either the Hamiltonian is not all that goes on, so there must be hidden variables taking place in a measurement process; or the Schrödinger equation is not all that goes on, so there must be another non-linear dynamic occurring in the measurement problem (e.g. a collapse); or what we experience is not all that there is. The latter solution is known as EQM: $|\psi\rangle$ expresses the universal wave function, and $|\psi_1\rangle$ and $|\psi_2\rangle$ represent relative states going on in different branches of reality. According to the general idea underlying EQM, there is a branching process whenever a quantum system enters into a superposed state ---being up for debate whether such a branching process is a branching of states or worlds.\footnote{We would like to thank an anonymous referee for pointing that out.}
In EQM, equation \ref{eq:simplesuperp} tells us that both cases are actual in two different states/worlds, so it tells us that each of its terms is the case:

\begin{itemize}
 \item The dead cat, with the poison flask broken, as a result of the falling atom, represented by $|\psi_1\rangle$;
 \item The living cat, with the poison flask intact, because of the atom that did not decay, represented by $|\psi_2\rangle$.
\end{itemize}

One of the challenges of EQM is to make sense of probabilities, insofar as every possible outcome does occur in a different branch. The most famous reaction to this problem is the \textit{fission program}, which basically says that we should leave it at that, insofar as objective probability does not make any sense in EQM at all \citep[see][]{greaves2007, tappenden2008}. Thus far, this is EQM. Enter its different readings of what is going on, with distinct ontologies, where the distinction between the relative facts interpretation (RFI)\footnote{We are adhering to Conroy's (\citeyear{Conroy2012everettnote, Conroy2018actualism}) terminology insofar we are discussing her work; nevertheless it should be noted that the traditional one is ``Relative-state interpretation''. \citet{Conroy2012everettnote, Conroy2018actualism} claims that the RFI is the closest interpretation of what \citet{everett1957relative} did, but we will not discuss that here.} and the many-worlds interpretation (MWI) is indispensable. They both use the same formalism to tell mathematical stories about state vectors and physical systems: Born Rule for probabilities and the Schrödinger equation to time evolution. As for the measurement problem, their response is also in tune with each other: no collapse. Here is where RFI and MWI begin to disagree: while the MWI says that the measurement problem is solved because the \textit{physical systems} somehow branch into distinguishable states, the RFI would say that what does branch into different states are the \textit{states} of physical systems.

Of course, EQM can be endorsed without any realist commitment at all. \citet{everett1957relative} himself ---in the reading offered by \citet{Barrett2011everettoneworld}--- supports an empiricist view on EQM (largely shared by Barrett). \citet{Conroy2012everettnote} does not differ substantially from Barrett's take on Everett. She calls Everett's view ``operationalist'', but ---as it will be discussed shortly--- she adds her own realist interpretation to what she takes to be Everett's own neutral stance on real ontology. It is worth pointing out the fact that the majority of Everettians today endorse many worlds realism (from David Wallace to Sean Carroll). There is a significant minority of Everettian empiricists in the tradition of Everett himself (for example Barrett), and there is a spectrum of other views, one of which is Conroy's RFI.

This is noteworthy because, while the mathematics of MWI is pretty much the same as RFI, their difference in terms of ontology is huge: by multiplying the physical systems into branches, the MWI multiplies \textit{whole physical worlds} ---hence ``\textit{many} worlds--- whereas the RFI points out that, since the superposition described in equation \ref{eq:simplesuperp} never fully breaks, it is strictly speaking not justified to speak of states confined to branches. The ontology for any object thus has to cover all branches. So, in terms of the accompanying intended ontology, we might obtain this:

\begin{itemize}
 \item[] RFI $\longrightarrow$ One world;
 \item[] MWI $\longrightarrow$ Many worlds.
\end{itemize}

This shows an ontological underdetermination within the EQM itself, i.e., one can ``read off'' from EQM information about a unique or about a plurality of worlds whose existence is `entailed' by the full functioning of physical theory.

The contemporary debate has extended to the nature of the modality, revolving around how theories about the modality could be scientifically informed by EQM. We find in the literature on the subject two theories recently put forward; both explain the nature of modality, taking EQM into account: Everettian actualism (EA), which is developed by \citet{Conroy2018actualism} from RFI and the quantum modal realism (QMR), which \citet{wilson2020QMmodal} builds upon MWI. While the former is an adaptation of Plantinga's \citeyearpar{plantinga1976} actualism, the latter builds from Lewis' \citeyearpar{lewis1986} modal realism, so the contemporary dispute between EA and QMR is a well-known one. Let us call it the ``Conroy--Wilson dispute''.

It is not our aim to criticize EA and QMR, but rather to point them out as suitable alternatives of naturalistic inclination for the understanding of modality based on EQM. Therefore, we will not do a full examination of the arguments, but we will focus on the key principles of both EA and QMR. Let us begin with the latter.

As \citet{wilson2020QMmodal} builds his QMR on the assumption of MWI, so we will survey its foundations. From this point onward, we will identify the MWI with the contemporary literature based on Wallace's (\citeyear{wallace2012emergent}) work, often called ``Oxford MWI'' or ``Oxford Everettians''. This cluster of proposals is roughly characterized by considering the branches as emergent phenomena from the universal wave function and dealing with probabilities through a decision-theoretic approach.

To the MWI, the terms of a superposition do not represent something real in the same world, but in different worlds that originate in the superposition described by equation \ref{eq:simplesuperp}. The notion of ``world'' in the MWI is defined by our own experience, which corresponds to the perspective of classical physics in the sense that there are no observable superpositions in a single world. For example, we perceive the state of the cat as always well defined, and we never perceive the other worlds. 

MWI provides an extra ingredient to the catalog of reality, which now contains parallel worlds. Thus, the ontology, which deals with existence questions, can be naturalized: we can indeed \textit{extract} from the MW the catalog of what exists (or at least part of it, given that one is being a realist about quantum mechanics as per many worlds). As \citet[p.~54]{Wallace:2011ib} puts it, taken at face value, MWI entails ontologically that ``we are living in a multiverse \textelp{} [MWI] claims that they exist, and so if the theory is worth taking seriously, we should take the branches seriously too. To belabour the point: According to our best current physics, branches are real.'' As physics, all that the MWI provides is the branching process. MWI tells us that the world branches out in concrete worlds in each superposition situation, so we are advised to include ``multiple worlds'' in our ontology ---in the sense of an item in the universe's catalog, according to the MWI. 

In the QMR terminology, each world is called an ``Everettian world''. The collection of all Everettian worlds is called the ``Everettian multiverse''. Our Everettian world i.e. the world in which you read this paper is not special: it is just one Everettian world among the infinitely vast Everettian multiverse, and objective chance (e.g. the Born Rule) indicates which Everettian world is ours \citep[p.~22]{wilson2020QMmodal}.

Necessity is taken to be a fundamental concept, so the nature of modality is to be explained. In QMR, this is done with MWI. While Lewis' (genuine) modal realism encompasses metaphysical, logical, and physical possibilities, Wilson's quantum modal realism reduces the Lewisian worlds into Everettian worlds, i.e., the existing worlds are only those arising from the physical possibilities spelled out from EQM in the context of the Schrödinger equation which describes the universal quantum system. As \citet[p.~22]{wilson2020QMmodal} describes, this is the principle of \textit{Alignment}: ``\textelp{} to be a metaphysically possible world is to be an Everett world''.

So, instead of attributing reality to \textit{every conceivable situation}, the Everettian world emerges in situations in which MWI's branching process occurs. For example, in the Schrödinger-cat-like situation, as described by Eq. \ref{eq:simplesuperp}, the physical situations described by the vectors $|\psi_1\rangle$ and $|\psi_2\rangle$ are both real in different worlds, i.e., a happy world in which the cat is found alive and a sad world in which the poor feline is found dead. But notice that the laws of EQM are maintained in every world, which are fundamental ``\textelp{} across the whole Everett multiverse'' \citep[p.~170]{wilson2020QMmodal}.

Notice: all these moves are done by \citet{wilson2020QMmodal} in order to explain \textit{modality} inspired in terms of the physical concepts of Everettian world and Everettian multiverse: the Everettian multiverse is metaphysically necessary, whereas each Everettian world comprises what is metaphysically possible; objective chance, through the laws of probability of EQM, enables us to identify what are the objective chances for events to occur in our Everettian world.

But the conclusion that to reify the existence of many concrete possible worlds is necessary to make sense of EQM is not undisputed. Conroy's (\citeyear[p.~33]{Conroy2018actualism}) RFI is a one-world interpretation of EQM, which takes the quantum-mechanical description to be ``\textelp{} factual and counterfactual descriptions of the world''. To make sense of the branching process, \citet{Conroy2018actualism} develops the EA, building upon Plantinga's actualism. EA considers that possible worlds are states of affairs: this is, as \citet[p.~30]{Conroy2018actualism} notes, ``the major difference between QMR and EA''.

The EA is straightforward. Each term in a superposition equation such as \ref{eq:simplesuperp} describes a possible world. They are all real, but only one world ``obtains'', that is only one term that describes our actual world. There is only one actual world, and we happen to live in it and have epistemic access to it. Other terms are just counterfactual descriptions. In Schrödinger's cat situation, EA would say that there is no absolute fact of the matter regarding the state of the poor cat. Rather, the result that the cat is ``alive'' is relative to the cat being ``dead'' and mutatis mutandis the result that the cat is ``dead'' is relative to the cat being ``alive''. EA is surely less ontologically extravagant than QMR in this sense, but the drawback is that it leaves modality unexplained. \citet{Conroy2018actualism} acknowledges this and argues that this is no prima facie reason to believe this cannot be done in future works, so it cannot be a sound argument to rule out EA. In disagreement, \citet[p.~95]{wilson2020QMmodal} states that ``\textelp{} abandoning Physical Actualism clears the way for a powerful and reductive theory of metaphysical modality''. The same goes for probability.

In terms of realism about worlds, the Conroy--Wilson dispute is a familiar one: it resembles the famous debate between David Lewis and Alvin Plantinga concerning modal realism and actualism. On the QMR site, the most adequate way of interpreting what EQM is telling us about reality is to adhere to the plurality of Everettian worlds as real as this one, with the same laws of EQM and so forth. On the EA side, this is not mandatory, and we can make sense of what EQM tells us about reality with a single-world theory. So there is an ontological underdetermination that will bother scientific realists of the shallow kind.

\label{page} Let us consider this issue more carefully. The matter is delicate, because ---in general--- it is assumed that the debate about possible worlds is, automatically, a debate about metaphysics qua metaphysics. This is because, as an anonymous reviewer carefully noted, \citet[6]{Conroy2018actualism} follows Planinga by saying that possible worlds are ``abstract sets of possible states of affairs'', and so there is only one concrete world. Wilson follows Lewis saying that they are concrete worlds, and that's why there are many. Thus, the ontological difference (the number of worlds that exist) depends on the metaphysical position as to the nature of these objects. Therefore, the two sides of the debate not only propose deep metaphysics but differ ---including--- in this deep part (and not only in the shallow-ontological part).

Our point is that, at least as the debate has been conducted so far, it is a debate of metaphysics \emph{qua ontology}. Take, for example, David Lewis's general plan, which is to quantify over worlds. On this plane, worlds are objects ---in the usual sense of object. It is a Quinean-style approach to ontology, so to speak, but assuming the worlds in the scope of the quantifiers--- which aims to obtain the explanatory power acquired when this is assumed. Alvin Plantinga's general plan is similar: to have the same explanatory power when talking about worlds, but paraphrasing. So it's also, in a way, a Quinean ontology strategy, in which someone is allowed to talk about certain things without having to assume them to exist in the ontology. The metaphysical question is different. In the case of objects: if there are objects, what is their metaphysical profile? For \citet{french2014}, \citet{branding2012OSRunderdet}, this is mainly asking about their individuality profile. The same can be said about worlds. If worlds are objects, they are in the scope of the existential quantifier. To say that they are concrete or not, emergent or fundamental, still seems little to give this metaphysical profile. Take again the case of objects and their metaphysical profile. One could say, for example, that there are individuals, and then point to the many individual tables and chairs that inhabit the world. This, however, still does not concern the metaphysical profile of individual tables and chairs. The individuality profile of tables and chairs should be specified in terms of e.g. theory of bundles of properties, such as tropes, or of object substrates, such as haecceities.

\section{The Dilemma}\label{null:sec:4}

So far, the whole dispute between QMR and EA has been stated as an ontological dispute concerning the existence of many worlds. What about the deep scientific realism for QMR and EA? The Conroy--Wilson dispute is an \textit{ontological} dispute: it concerns the existence of the worlds in EQM. According to the demands imposed by the Chakravartty Challenge, it remains at the shallow level: no metaphysical description of the nature of the worlds is put forward. As an anonymous referee pointed out, this might seem unfair, as ``these worlds come with specific properties, and the mechanisms for the splitting of worlds are specified in several papers on MWI. Thus, one has some notions about the features of the Everettian worlds''. However, as we argue, the properties of the world specified in the huge literature on the MWI and RFI is mainly a description in physical and ontological terms, not on metaphysical terms, viz. the metaphysical qua metaphysical profile of the worlds. So, none of these addresses yet the kind of metaphysical question that deep realism requires, i.e., addressing those questions concerning the existence of  worlds (an ontological question) is still a matter of `reading the equations', not of diving into deep metaphysics (the `nature' question). The deep realist adhering to the Toolbox methodology will certainly complain: shallow scientific realism is not realist enough. We need a metaphysics attached to it. Given that none seems to be found yet, what are we supposed to do with EA and QMR? 

Recall, as we saw in section \ref{null:sec:2}, that \citet[p.~48]{french2014} accused shallow realists (viz., the ones that do not recognize the need for a metaphysical image over the ontology) of being closet empiricists or `ersatz' realists. It would seem that we are being recommended not to embrace these interpretations if we want to be realists, given that they are not deep enough. However, if we do not allow that this `fleshing out' be framed in terms of an interpretation only (given that this only delivers the ontology), but must go somehow deeper, then a dilemma is posed: either we may embrace EA and QMR as legitimate forms of realism about QM, or else we reject them on the basis of being shallow.

To our best knowledge, there is no textual indication showing that the defenders of deep realism take specifically the considerations EA and QMR to be exemplifications of shallow realism \citep[cf., for instance, ][]{french2020}; however, as shown in section \ref{null:sec:2}, if shallow realism is to be essentially restricted to existence questions ---which we take to be the crux of the matter of the dispute between actualism and modal realism\footnote{Recall, again, that the debate between Lewis and Plantiga is a dispute about the furniture of the world!}--- then such views could be reasonably framed as `shallow'. On the other hand, however, one is left with no clue on how to go `deep' in these cases. Can we choose some of the horns of the dilemma and argue for our choice? The idea that QMR and EA \textit{are} legitimate forms of scientific realism ---viz. about quantum mechanics--- has considerable pedigree.
It can certainly be argued that interpreting quantum mechanics is naturalistically well-grounded and even advised, given that the demand for an interpretation \textit{comes from science itself}:

\begin{quote}
    \textelp{} it is crucial to observe that the demand for interpretation arises within theoretical science. It isn't something imposed on the scientist by some aprioristic philosopher. Yes, scientists themselves may have been corrupted by exposure to the specious demands of philosophy and imported this unnecessary puzzlement into their views on their science. \citep[p.~1123]{sklar2010naturalist}.
\end{quote}

So, in this sense, ontology has close ties to the science one is concerned with. \citet{esfeld2019measurement, maudlin2019quantumphil} go even further, arguing that such an interpretative demand is essential for us to have a proper \emph{physical} theory, which should be composed of: (i) the specification of a dynamic law, which spells how things evolve through space in time, and (ii) an ontology, which says what the theory is about. So, repeating physics does involve considering an interpretation, and this already goes a long step toward addressing the worries raised by French. 

It could be further argued that French has given a somewhat distorted image of the shallow version of realism. That could be seen more clearly if we notice that for some, realism, in a naturalistic sense, is just that: repeating what science says, \emph{interpretation included}. In particular, one would not even need to add more `philosophical' aspects, such as claims to the effect that the theories are coming closer to the ultimate truth. So, in this sense of realism, by definition, one has nothing better to do than just repeat science, and the addition of a metaphysical layer seems to sin against realism, by adding unnecessary epistemic risk. Consider Burgess:

\begin{quote}
For many professed `realists', realism amounts to little more than a willingness to repeat in one's philosophical moments what one says in one's scientific moments, not taking it back, explaining it away, or otherwise apologizing for it: what we say in our scientific moments is \textit{all right}, though no claim is made that it is \textit{uniquely} right, or that other intelligent beings who conceptualized the world differently from us would necessarily be getting something wrong. For many professed `anti-realists', realism seems rather to amount to a claim that what one says to oneself in scientific moments when one tries to understand the universe corresponds to Ultimate Metaphysical Reality, that it is, so to speak, a repetition of just what God was saying to Himself when He was creating the universe. \citep[p.~19, original emphasis]{burgess04}.
\end{quote}

So, in favor of shallow realism, one can advance a good amount of considerations: interpreting is part of what science itself demands, and keeping oneself restricted to what science says is part of what realism itself, at least in a naturalistic version, requires. If that is correct, the pressure now is on the deep realist, who urges us to answer Chakravartty's Challenge if we are to be legitimate realists, and it is incumbent on her to provide reasons recommending that we go deeper than that. Notice that merely saying that one cannot repeat what science says is too little to move the shallow realist, because that is precisely what the shallow realist is willing to do.

As a result, if the deep realist is not going to beg any question against the shallow realist, perhaps we could turn the table and ask what are the advantages of adding a metaphysical layer neither required nor endorsed by science. Notice: if the deep realist can grant her point, then, we must agree that EA and QMR are not proper versions of realism; that is quite a bitter pill, but some may be willing to swallow it. So, we are right in asking the deep realist for a nice cup of water. Let us check whether some relief can be found. 

One of the alleged roles of metaphysics coming from the toolbox from analytical \textit{a priori} metaphysics concerns aid to understanding science \citep[see][]{French2018RealMetaph,french2018jgps}. Given that contemporary physics is so far away from our common-sense picture of the world, adding a metaphysical layer that is connected to such an intuitive picture may provide some of the sought understanding of contemporary science. Applied to the case of EA and QMR, the point in question would require that we do not clearly understand the picture of reality that these interpretations are providing for us, and by not having this connection to common sense via metaphysics, the views are not properly intelligible, or, perhaps, deliver not a proper understanding, and rank poorly when it comes to alternatives. 

But notice that this demand \citep[p.~228]{french2018jgps} of metaphysical ingredients tied with a previous non-scientific view for providing understanding seems to be at odds with the very role that interpretations are thought to fulfill. An interpretation is part of an attempt to provide an image of the world according to the theory, and it certainly delivers more than merely reciting again the equations of the theory. So, the demand that we go deeper than the ontology (provided by the interpretations) seems to be based on a distortion of the image of reality that results without an embrace of metaphysics.

There is also an additional worry. The plan suggested by French is that deep realism will be able to advance a metaphysical image of science, grounded in a common-sense picture that comes along with the metaphysical content \citep[p.~228]{french2018jgps}; this image will allow one to make sense of physics by this connection of the metaphysics with a more familiar image of reality. But now the worry is that such attempts may end up very much performing the role of domesticating physics, as largely criticized by \citet{Ladyman2007evmustgo}. If that is the case, then, again, we do not end up with a clear picture, but with a distortion of science.

Connected with this point, one could add a further difficulty: if science needs metaphysics to be intelligible, and realistic versions of quantum mechanics such as EA and QMR are shallow, and, therefore, should not be embraced because they fail in being clear enough according to \textit{those} standards, then, there is the suspicion that we may end up judging science from the point of view of the demands of a first philosophy, which stipulates from the armchair the bounds of scientific intelligibility. Now, that is a problem even in the context of the Toolbox of metaphysics, where metaphysics is at the service of science, and not the other way around. It seems that this is a \textit{modus tollens} against the very idea that Chakravartty's Challenge regulates what is a legitimate form of realism, given that it places science below metaphysics.\footnote{Notice that despite the name of the Challenge, \citet[p.~160]{chakravartty2017scientific} clearly does not endorse it (at least in the sense put forward by French): ``[i]t is simply not the case that every last question regarding the ontological natures of things must be settled in order for realism about them to be viable!''. This was also called the ``Meta-Chakravartty's Challenge'' \citep{arroyoarenhart2021}. See also \citet{chakravartty2019challenge} for more details about his denial of his own Challenge, as framed by \citet{french2014}.} 

All of this notwithstanding, suppose the deep realists still want their clear picture about EQM: what is such a clear picture that we are looking for, after all? In order to grasp what is at stake with such a demand, let us recall that the deep realism was proposed by \citet{french2014} in the context of the individuality of quantum objects: the \textit{deep} part consists in the attribution of what \citet{branding2012OSRunderdet} called the ``individuality profile'' of quantum objects with the latter being the ontology of the theory. As it is known, such a metaphysical profile concerning individuality cannot be scientifically chosen, due to metaphysical underdetermination \citep[for the details of this debate, see][chap.~4]{KraFre2006}. However, in this debate we know what we are looking for: the metaphysics of individuality ---are quantum objects individuals or non-individuals? \citep[see][]{arenhart2017receivedview}. To be sure: ``\textelp{} when the object-oriented realist is unable to tell us whether electrons, qua objects, are individuals or not, it risks vacuity'' \citep[194]{french2016}. But the same cannot be said about the case of worlds: what is the kind of answer that deep realists are looking for when asking about the \emph{metaphysical profile} of possible worlds? What are the options? We just don't know. And that's a methodological drawback to deep realism, given that we can only look for answers when we know the questions.

Finally, we would like to point to an additional methodological problem for deep realism: how deep is deep \textit{enough}? As pointed out by \citet[p.~12, original emphasis]{chakravartty2019}, in metaphysics ``[i]t is \textit{always} possible to ask finer-grained questions \textelp{}''. In fact, this is why the analogy of a ``slippery slope'' is suggested \citep[cf.][]{chackravartty2013}: at the risk of falling down the slope, one should not demand that all metaphysical specifications be detailed in order to point for a fruitful relationship between metaphysics and science ---as is the motivation of deep scientific realists--- but, on the contrary, we must follow a metric according to which proximity to science would ensure that we cross the slope safely, viz. without falling into endless metaphysical speculations. However, this same metric only poses the methodological dilemma in other words: for, as we have seen, too close to science would leave us on the shallow side of realism, while complete detachment would leave metaphysical investigations completely adrift of science. That said, the additional problem, which we could call the ``threshold problem'', is the (until now) lack of perspectives to justify how much metaphysics one must put into scientific realism for it to be considered deep enough. In other words, the threshold problem indicates a dangerous vagueness in the notion of depth: should it be neck-deep or Mariana-Trench deep? This issue remains unaddressed.

So, let us summarize what the deep realist has to face: 1) the demand for a deeper level including metaphysics begs the question against the shallow realist; 2) the idea that without metaphysics our image of science is merely a reciting of the physics, with no clear image, is largely exaggerated; interpretations of quantum mechanics, such as EA and QMR are precisely attempting to fill a gap in providing an image of the world, and are, besides, scientifically motivated; 3) this trumps the positive account of the deep realist, according to which adding metaphysics is a must because only by providing for such addition we can understand science. The latter, besides not being motivated, faces its own difficulties, because 4) it leads us to a domesticated view of science, and to judge science from the point of view of metaphysics; and 5) even if one adheres to deep realism, it is not clear what should one look for in the case of worlds, i.e., it is not clear what can be considered a metaphysical profile for possible worlds and how such profiles may add to our understanding of what a world is.

In this way, deep realism tells us where to look for the metaphysical layer in the case of an ontology of objects. We do not know, however, which tools are to be sought in the metaphysics toolbox for ontologies in which objects are possible worlds ---whether they are understood as concrete à la QMR or possible à la EA. Even more: specifically, postulating worlds in quantum mechanics serves the purposes of quantum mechanics, in the same way that positing electrons in quantum mechanics serves the purposes of quantum mechanics ---and not metaphysics (qua ontology) in general. So one might even argue that the relevant ``tools'' in the analytic metaphysics toolbox, as being employed by QMR and EA, are not metaphysical tools qua metaphysics, but \textit{qua ontology}\footnote{Recall that we have defined in section \ref{null:sec:2} ``ontology'' as a subset or branch of ``metaphysics''.} \citep[among examples of such a conflation are Wilson's book reviews, e.g.][]{french2020, arroyo2021}: metaphysics still, but not of the deep kind that deep realists were demanding all along. One is still reading the theory, not looking at the metaphysician's toolbox. That means that we are stuck at the level of ontology. Of course, there is the issue of whether such a postulation of worlds is the best option in quantum mechanics, but that is another question.

Our central objection, therefore, is in relation to the deep realism demand that the QMR and EA proposals are not considered sufficiently realistic proposals until such a metaphysical profile about worlds is specified. On the contrary, they seem to us to be legitimately realistic proposals despite filling the metaphysical gap demanded by deep realism. First because, in the case of possible worlds, this is vague: we don't know where to start; secondly, because the demand itself is very onerous.

In this sense, perhaps, what requires justification is the demand that metaphysics can and should act as an extra layer over science, as a tool employed to enlighten a scientific theory. But this, alas, is the hot topic of justifying metaphysics by relating it to science. And with that topic, we shall not be concerned now. 

\section{Concluding remarks}\label{null:sec:5}

In this paper, we investigated the demand that one needs metaphysics to inform scientific realism, the so-called project of `deep' scientific realism. We used EQM as a case study, and two of its interpretations concerning the existence of many worlds: QMR and EA. We found that the debate concerning QMR and EA is, as the old debate between modal realism and other accounts of `worlds', a debate upon the modes of existence (or non-existence) of such worlds. This is not a metaphysical debate \textit{qua} metaphysics, but metaphysics \textit{qua} ontology, i.e., concerning their \textit{existence}, not their \textit{nature}. This is as far as scientific realists of the shallow sort would go. So, in a nutshell: are we deep realists about EQM thus far? No. Are we entitled to go deep into metaphysics? Well, it seems that nothing is stopping anyone to delve into deep metaphysical questions in order to metaphysically inform our current best science, so yes. But \textit{should} we? We are not sure about that.

We must not legislate that legitimate realism is exclusively that form of realism informed by deep metaphysical questions. To begin with, because this is vague ---viz. how deep should someone go in the pool of metaphysics? But mainly because scientific realism is an attitude that is more interested in the relationship between theory and the world, so it cannot depend on the (fragile) justification of metaphysics as a discipline to be able to legitimize itself as a meta-scientific attitude.

\section*{Conflict of interest}
The authors declare that they have no conflict of interest.

\section*{Data availability}
Data sharing not applicable to this article as no datasets were generated or analysed during the current study.
\bibliographystyle{spbasic}
\bibliography{foop.bib}
\end{document}